\documentclass[conference]{IEEEtran}
\IEEEoverridecommandlockouts
\usepackage{amsmath,amssymb,amsfonts}
\usepackage{algorithmic}
\usepackage{graphicx}
\usepackage{textcomp}
\usepackage{xcolor}
\usepackage{todonotes}
\usepackage{tabularx, multirow}
\usepackage{fancyhdr,lipsum}

\def\BibTeX{{\rm B\kern-.05em{\sc i\kern-.025em b}\kern-.08em
    T\kern-.1667em\lower.7ex\hbox{E}\kern-.125emX}}
\usepackage[%
backend=biber,bibencoding=utf8, %instead of bibtex
language=auto,
style=ieee,
sorting=none, % nyt for name, year, title
maxbibnames=10, % default: 3, et al.
natbib=true % natbib compatibility mode (\citep and \citet still work)
]{biblatex}
\bibliography{main.bib}
\def\app#1#2{%
  \mathrel{%
    \setbox0=\hbox{$#1\sim$}%
    \setbox2=\hbox{%
      \rlap{\hbox{$#1\propto$}}%
      \lower1.1\ht0\box0%
    }%
    \raise0.25\ht2\box2%
  }%
}
\def\approxprop{\mathpalette\app\relax}

\graphicspath{{figures/}}

\begin{document}
\title{Enhancing the German Transmission Grid Through Dynamic Line Rating
\thanks{Fabian Hofmann is funded by the Breakthrough Energy Project ``Hydrogen Integration and Carbon Management in Energy System Models''.}
}

\author{\IEEEauthorblockN{1\textsuperscript{st} Philipp Glaum}
    \IEEEauthorblockA{\textit{Institute of Energy Technology} \\
        \textit{Technical University of Berlin}\\
        Berlin, Germany \\
        p.glaum@tu-berlin.de}
    \and
    \IEEEauthorblockN{2\textsuperscript{nd} Fabian Hofmann}
    \IEEEauthorblockA{\textit{Institute of Energy Technology} \\
        \textit{Technical University of Berlin}\\
        Berlin, Germany \\
        m.hofmann@tu-berlin.de}
}

\maketitle

\begin{abstract}
    The German government recently announced that 80\% of the power supply should come from renewable energy by 2030. One key task lies in reorganizing the transmission system such that power can be transported from sites with good renewable potentials to the load centers. Dynamic Line Rating (DLR), which allows the dynamic calculation of transmission line capacities based on prevailing weather conditions rather than conservative invariant ratings, offers the potential to exploit existing grid capacities better. In this paper, we analyze the effect of DLR on behalf of a detailed power system model of Germany including all of today's extra high voltage transmission lines and substations. The evolving synergies between DLR and an increased wind power generation lead to savings of around 400 million euro per year in the short term and 900 million per year in a scenario for 2030.
\end{abstract}


\section{Introduction}
The deployment of new renewable energy sources far from load centers is leading to an increasing need for grid capacity. A notable example can be found in Germany, where most of the wind energy is generated in the North while there is significant industrial demand in the South \cite{kuhneConflictsNegotiationProcesses2018}. The current federal government plans to add ~70 GW wind power capacity until 2030, at the same time the transmission line capacities need to be expanded \cite{federalministryofeconomicaffairsandclimateactionGermanyCurrentClimate2022}. However, in the last decade only few grid expansion projects have been realized. Most of these are delayed due to administrative problems and protest activities \cite{neukirchProtestsGermanElectricity2016}, \cite{fernandezReviewDynamicLine2016}. This trend will likely exacerbate transmission system congestion, which may lead to more grid instability and curtailment \cite{erdincComprehensiveOverviewDynamic2020}. There is therefore a strong incentive to make better use of the existing grid infrastructure.

To overcome the mismatch between rapid installation of renewable generation and slow installation of transmission lines, several, complementing measures can be taken that must be reconciled:  (1) large scale implementation of storage facilities to flatten power feed-ins and power demands \cite{caoPreventPromoteGrid2021}; (2) usage of alternative energy carrier networks like hydrogen to relieve the electricity grid \cite{welderDesignEvaluationHydrogen2019}; (3) leveraging the existing grid infrastructure to increase the operating capacity. In the following, we will focus on the last point, which in comparison to (1) and (2) stands out through a fast, low-cost and non-invasive implementation \cite{deutscheenergie-agenturgmbhdenaErgebnispapierDenaStakeholderprozessHohere, minguezApplicationDigitalElevation2022}.

Traditionally, transmission line capacities are calculated assuming unfavorable, static weather conditions such as 40$^\circ$C  ambient temperature and 0.6~m/s wind speed \cite{fernandezReviewDynamicLine2016}. This is referred to as Static Line Rating (SLR). By design, SLR underestimates the capacity of a transmission line and, when implemented in practice, leads to an underutilization of the transmission infrastructure. In contrast, Dynamic Line Rating (DLR) calculates the line capacity taking into account the prevailing weather conditions. Cold weather and wind cool overheated transmission lines, enabling the thermal rating to be raised. This results in key benefits in cost-efficiency, congestion reduction and better wind power integration \cite{erdincComprehensiveOverviewDynamic2020}.
Transmission System Operators often adjust line ratings based on the current season but do not exploit the full potential of DLR monitoring local cooling effects \cite{bundesnetzagenturMonitoringReport20212022}.
The literature provides case studies with DLR applied in small scale Energy System Optimization Models (ESOMs) \cite{liDayAheadSchedulingPower2019,dabbaghjamaneshEffectiveSchedulingReconfigurable2019,viaforaDayaheadDispatchOptimization2019}. %Rena suggested a reference. However, is a simulation and not an optimisation, would not put this reference here. Results not comparable as different energy system.
% IEEE paper Increase cross-border capacity to reduce market splitting of day-ahead electricity markets A dynamic line rating approach; Consentec paper: https://www.currenteurope.eu/wp-content/uploads/2021/12/currENT_Consentec_BenefitsOfInnovativeGridTechnologies_FinalReport_20211208_clean.pdf
However, it is lacking of an extensive evaluation and assessment of DLR in higher scale ESOMs.
This paper addresses this gap. We present the first capacity optimization of the German power system with DLR being subject to high CO2 emission targets.

The article is structured as follows.
Section~\ref{sec:methodology} describes the methodology of the DLR implementation (\ref{sec:dlr}) and the underlying power system modelling (\ref{sec:powersystem}). In Section~\ref{sec:study-cases}, we present two case studies for the German power system used throughout the paper. Section~\ref{sec:results} comprises the main results of the analysis with limitations outlined in Section~\ref{sec:limitations}. A conclusion is presented in Section~\ref{sec:conclusion}.

% The underlying model is created from the PyPSA-Eur framework [16], and for Germany consists of 256 representative nodes and considers a time horizon of one year with an hourly resolution. To model the effect of DLR on the power system, we use the IEEE standard \cite{IEEEStandardCalculating2012} with highly resolved historical weather data implemented in the Atlite software \cite{hofmannAtliteLightweightPython2021}. The advantages of the DLR are illustrated by comparison with an SLR-based system. Therefore, we analyze and compare curtailment, grid congestion as well as the optimal generator deployment.

\section{Methodology}
\label{sec:methodology}

\subsection{Dynamic Line Rating}
\label{sec:dlr}

For simulating the effect of DLR, we follow the IEEE standard documented in \cite{IEEEStandardCalculating2012}, \cite{IEEEStd738}.
The key concept of DLR is based on a dynamic estimation of the maximally allowed electrical current  for the conducting material. It is set such that the conductor does not surpass the maximally allowed temperature  after which impermissible sag of the line or hardware damage can be expected \cite{KARIMI2018600}. To model the weather dependent impact on DLR, we implemented the IEEE standard in Atlite \cite{hofmannAtliteLightweightPython2021}, a Python package used for converting weather data into renewable power potentials. The following outlines the basic concept of DLR, for further detail refer to \cite{IEEEStandardCalculating2012}, \cite{IEEEStd738}.
For each conductor, the heat balance equation
\begin{align}
    q_c +q_r = q_s + I^2 \cdot R(T)
    \label{eq:heat_balance}
\end{align}

relates the heat losses on the left hand side to the heat gains on the right hand side. Convective heat loss $q_c$, which represents cooling by ambient air, depends on ambient temperature, wind speed and angle, conductor material and geometry. The radiated heat loss $q_r$ is the net energy lost through black body radiation. Solar heat gain $q_s$, on the other hand, is caused by solar heat radiating  onto the conductor. Finally, the resistive heat gain $I^2 \cdot R(T)$ is given for an electrical current $I$ and temperature-dependent resistance $R(T)$, where $T$ is the temperature of the conductor.  The latter can be approximated by linearly extrapolating from reference resistance $R_{ref} = R(T_{ref})$ using a material specific temperature coefficient $\alpha$, i.e.
\begin{align}
    R(T) = R_{ref} \cdot (1 + \alpha \cdot (T - T_{ref}))
\end{align}

\begin{figure}
    \centering
    \includegraphics[width=\linewidth]{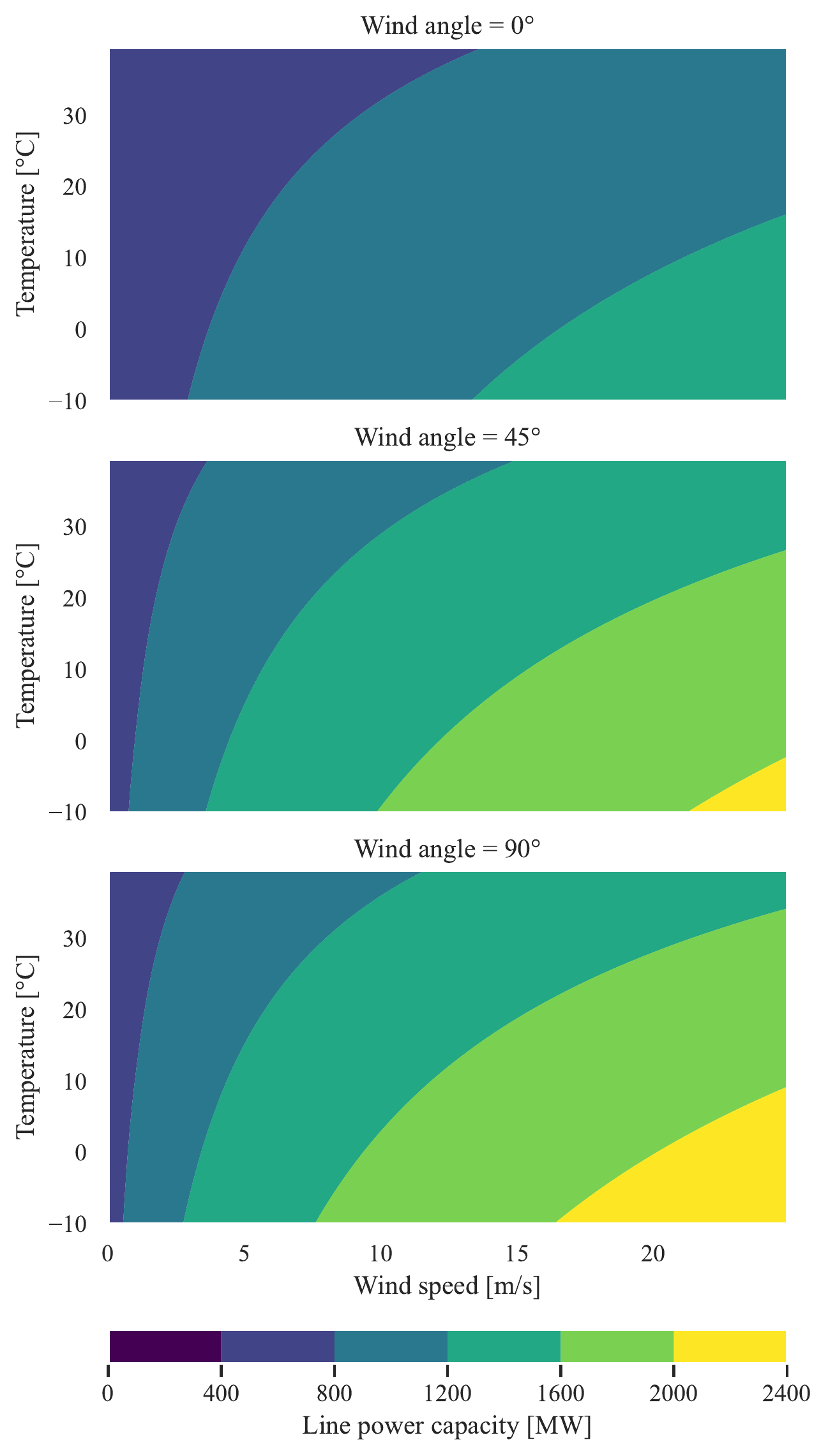}
    \caption{Transmission capacity for different environmental conditions. For three different wind incidence angles (0$^\circ$, 45$^\circ$, 90$^\circ$), the maximal transmission capacity of a typical electricity line ($R(T_{max})$=$9.39\cdot10^{-5}\,\Omega/\text{m}$, $V$~=~380~kV) is shown as a function of temperature and wind speed. The cooling effect of a perpendicular, strong, cold wind can lead to capacity increase of factor 4-5 compared to conservative conditions with 40$^\circ$C and low wind.}
    \label{fig:parameter-space-reduced}
\end{figure}

Solving \eqref{eq:heat_balance} for the electrical current and setting the temperature to its maximally allowed limit $T_{max}$, yields the ampacity
\begin{align}\label{ampacity}
    I_{max} = \sqrt{\frac{q_c +q_r-q_s}{R(T_{max})}}
\end{align}
which for three-phase electric power transmission operating at voltage level $V$ leads to a maximally allowed, constant power transfer of
\begin{align}
    P_{max} = \sqrt{3} \cdot I_{max} \cdot V
\end{align}
Fig.~\ref{fig:parameter-space-reduced} shows $P_{max}$ for a single wire of a typical 3-phase transmission line with $R(T_{max}=80^\circ C) = 9.39\cdot10^{-5}\,\Omega/\text{m}$ and $V$~=~380~kV, as a function of temperature and wind speed for three different wind incidence angles, 0$^\circ$, 45$^\circ$ and 90$^\circ$. The cooling effect at cold temperature with strong perpendicular wind leads to a transmission capacity increase of factor 4-5 compared to conservative conditions with 40$^\circ$C and low wind.
\subsection{Power System Modeling}
\label{sec:powersystem}
Using the Python package \textit{Python for Power System Analysis} (PyPSA) \cite{brownPyPSAPythonPower2018,browntomPyPSAPythonPower2022}, the linear ESOM is represented by a set of buses which are interconnected via transmission lines and complemented by loads, generators and storage facilities. Further, time-dependent electric demand per bus and generation potentials per generator are included for one year with hourly resolution. The generator dispatch and deployment is determined by minimizing the total system cost, where the optimization uses the linearized power flow approximation \cite{purchalaUsefulnessDCPower2005,brownLinearOptimalPower2016} and may be subject to systemic constraints (CO$_2$ budget, limited capacity expansion, etc.).

Data on the transmission grid, power plants, renewable potentials and demand are created with the workflow \textit{PyPSA-EUR} \cite{horschPyPSAEurOpenOptimisation2018,horschjonasPyPSAEurOpenOptimisation2022}. Underlying weather data is provided by the ECMWF Reanalysis v5 (ERA5) dataset providing various weather-related variables in a hourly resolution on a $0.25^\circ \times 0.25^\circ$ grid. The representative network of Germany comprises all 256~substations and 333~transmission lines operating at 220~kV and above. Electrical parameters of transmission lines are derived by mapping voltage level of the lines to standard line types given in \cite{oedingElektrischeKraftwerkeUnd2011}. %\todo[inline]{add descripion for geographical overlap and mapping of weather data.}
For DLR scenarios, the transmission capacity per line and time step is given by the formulation in Section~\ref{sec:dlr}, for SLR scenarios by the standard static transmission capacities. In case multiple weather grid cells overlap with a transmission line, the grid cell with the most unfavorable condition is chosen to provide the input variables. As the DLR scenario is calculated with averaged hourly wind speed data, sub-hourly wind speed fluctuations are neglected. This leads to a slightly overestimated $P_{max}$, which we consequently scale down by an empirically derived factor of $0.95$, see Appendix~\ref{dlr_factor} for details. Furthermore, we account for N-1 network security in both SLR and DLR scenarios by restricting the power transmission P per line to $-0.7\,P_{max} \le P \le 0.7\,P_{max}$ leaving a 30\% capacity buffer.
% \todo[inline]{add citation} https://www.50hertz.com/Netzlast/Karte/index.html also different, as they just categorize <70% as high utilization and we do it for N-1 security. Could not find a reference for this value.
Finally, since the error of the linearized power flow approximation increases with the voltage angle differences in the system \cite{dvijothamErrorBoundsDC2016}, we restrict the voltage angle difference $\Delta\theta$ across a line to maximally 30$^\circ$ by introducing the constraint
\begin{align}
	|P| \le \frac{V^2\Delta\theta}{x}=\frac{V^2\pi}{6\,x}
    \label{eq:angle_difference_constraint}
\end{align}
for every time step and line, where $x$ is the series reactance.

\section{Study Cases}
\label{sec:study-cases}
In the following, we present two study cases. These simulate the year 2019 for short-term benefits (i.e. by running an operational optimization) and 2030 for long-term benefits (i.e. by running a capacity expansion optimization). Both years are modelled using DLR and SLR. The parameters settings for both study cases can be found in the Appendix~\ref{tab:cases_parameters}.

\subsection{Optimal Operation of the 2019 Power System}
To quantify the effect of DLR on the existing power system, we choose the pre-pandemic year 2019 in which the system operated under relatively normal circumstances. Electrical load and renewable potentials are derived from historical data of the year 2019 \cite{muehlenpfordtTimeSeries2020}, transmission and generator infrastructure are aligned to the state of 2019, using selected data from \cite{hofmannFRESNAPowerplantmatchingV02022,openpowersystemdataDataPackageNational2017,wiegmansGridkitExtractEntsoE2016}. To approximate the operation of nuclear and lignite power plants mostly running at base load, we enforce a minimal requirement of operation, based on historical data from the ENTSO-E Transparency Platform  \cite{entso-eENTSOETransparencyPlatform2020}.

\subsection{Optimal Investment and Operation of the 2030 Power System }
In contrast to the 2019 scenario, the 2030 scenario of the German power system allows capacity expansion of renewable power plants, gas turbines, batteries and hydrogen infrastructure. From the existing power plant fleet, only those with a decommissioning date later than 2030 are included. Further, grid expansion projects from the TYNDP \cite{entsoeTYNDP2018Project2018}, which are reported to be built by 2030, are added to the grid infrastructure. In order to align the scenario with the goal of the federal government, 80\% of the electricity supply has to come from wind, solar and hydro power plants \cite{federalministryofeconomicaffairsandclimateactionGermanyCurrentClimate2022}. The electrical load time series, originally representing the base year 2013 of PyPSA-Eur \cite{horschjonasPyPSAEurOpenOptimisation2022}, is scaled up to meet the total predicted demand for 2030 \cite{kemmlerEntwicklungBruttostromverbrauchsBis2021}.

\section{Results}
\label{sec:results}

\begin{figure}[htpb]
    \centering
    \includegraphics[width=\linewidth]{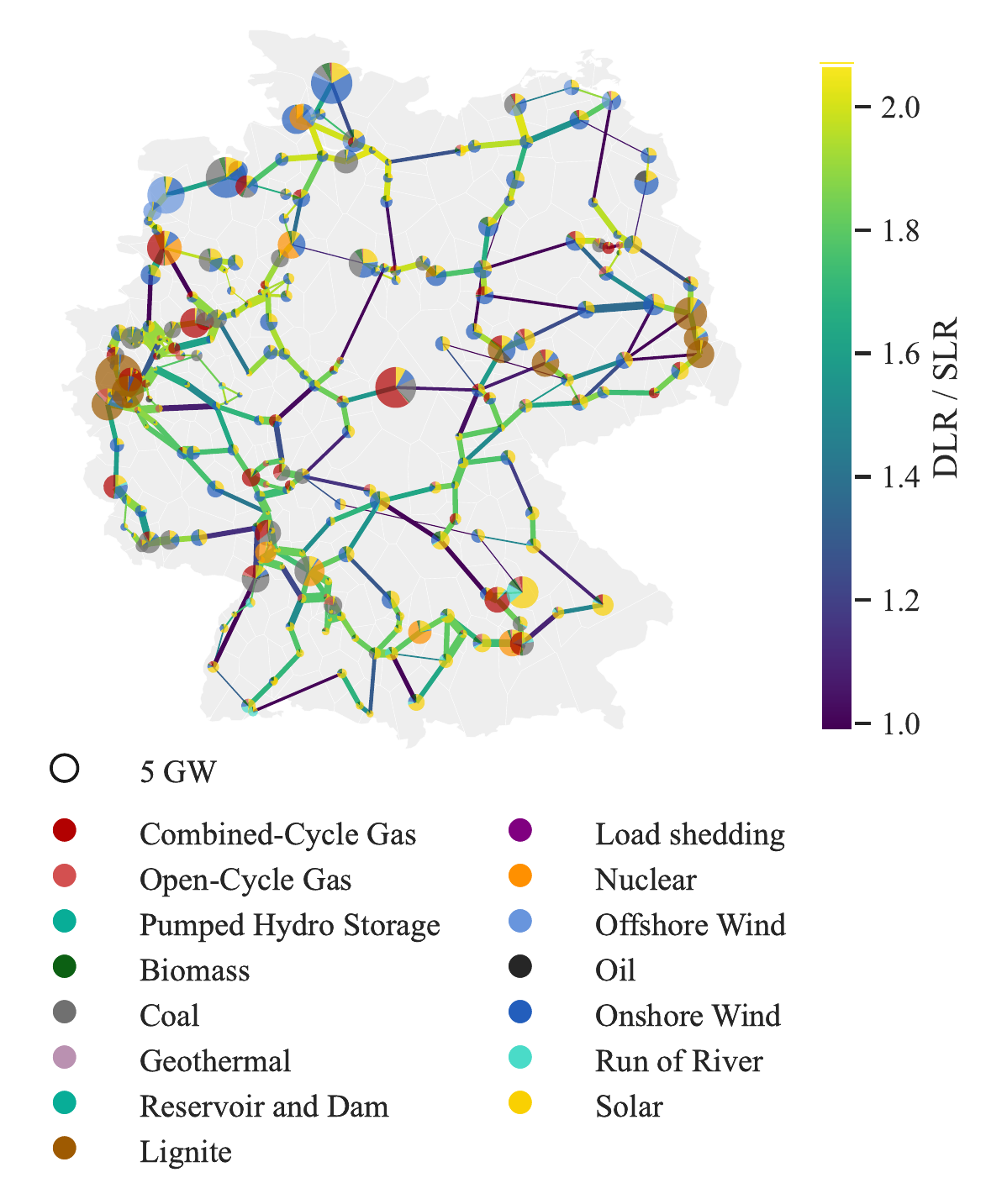}
    \caption{Capacity layout of the 2019 scenario of the German power system. The circle with its subdivisions are proportionally to the installed generation capacity. The widths of the line indicate their nominal transmission capacity, their color the relative average change when going from SLR to DLR.}
    \label{fig:capacity-map-2019}
\end{figure}

At first, we present the extent to which the DLR implementation alters the operation of the existing power system. Fig.~\ref{fig:capacity-map-2019} shows the geographical layout of existing generation and transmission capacities for the 2019 scenario. The circles with its subdivisions show the installed power generation capacity per site and technology with their area being proportional to the capacity. The widths of the transmission lines indicate the installed nominal capacities, while their colors show the relative average change in capacity when applying DLR. Note that for some of the long transmission lines with high reactances, the voltage angle constraint introduced in \eqref{eq:angle_difference_constraint} suppresses the relative improvement through DLR. For most of the others, an improvement in the average capacity up to a factor 2 is observed.
In particular, this accounts for short lines in the north and west of Germany.

\begin{figure}[htpb]
    \centering
    \includegraphics[width=\linewidth]{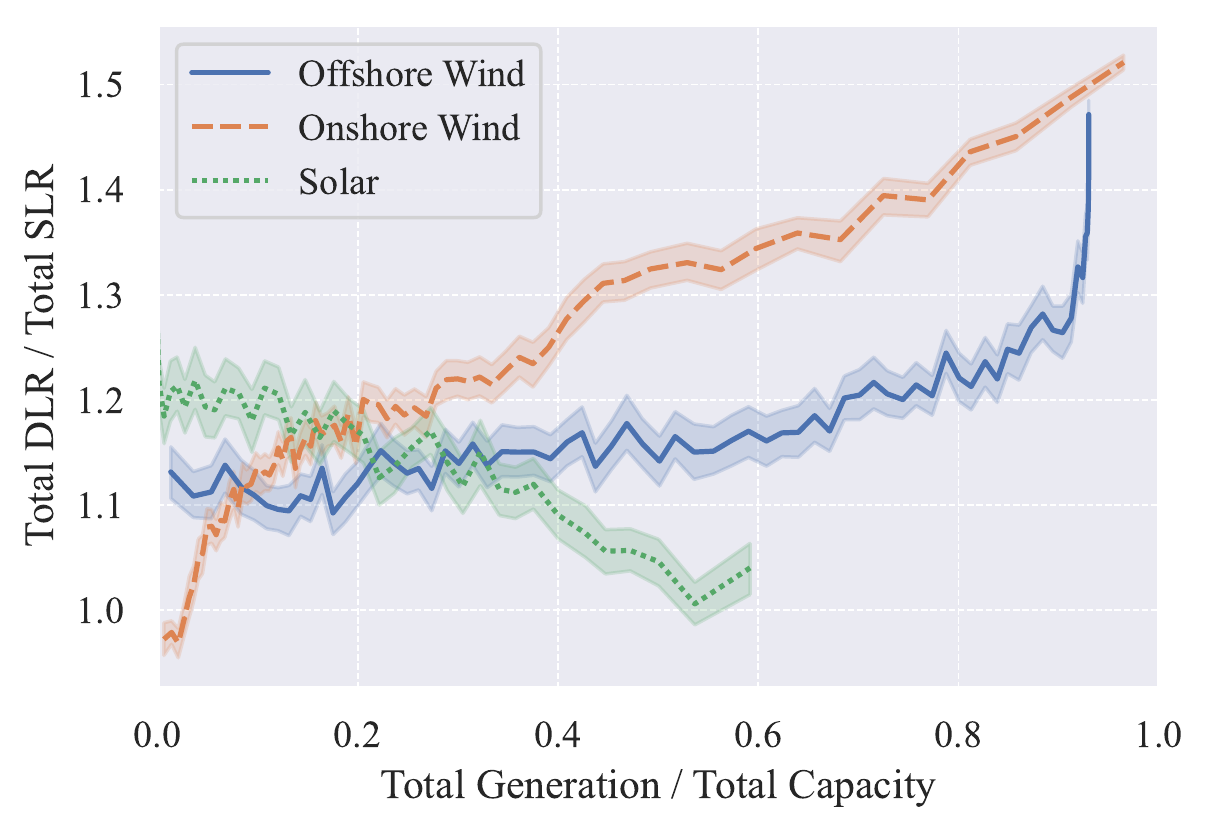}
    \caption{Ratio between total DLR and SLR transmission capacity as a function of the capacity normalized total generation. }
    \label{fig:potential-correlation-2019}
\end{figure}

In addition to location, it is of particular interest in what time periods the improvements occur. Fig.~\ref{fig:potential-correlation-2019} shows the total relative increase in transmission capacity as a function of the production relative to the installed capacity for solar, onshore and offshore wind separately. The lines represent the average when collecting the data points along the x-axis in groups of 30, with the shaded areas corresponding to the 95\% confidence intervals. The transmission capacity increase strongly correlates with the availability of onshore wind power in the system. In times when onshore wind power availability is at its maximum, i.e. close to 1, the transmission capacity increases by 52\% in average. For low offshore wind power potentials, the increase in transmission capacity weakly correlates with the availability. However, after a significant positive trend, 48\% transmission capacity increase is observed for times with high offshore wind availability. For solar power, the trend is roughly the opposite. In time periods with low solar potentials, the increase in transmission capacity averages to 20\%, while for a higher potential, the transmission increase declines. However, note  that for solar the overall transmission capacity of DLR stays above transmission capacity of SLR. Only at the very end, a strong positive trend is observed which originates from an exceptionally sunny and windy period in the end of April. %\todo[inline]{frame more positive, values are above 1 and solar mainly locally distributed.}

When running the optimization for the 2019 case study, these effects prove themselves to be quiet impactful. Of the 490~TWh total net electricity generation throughout the year, 31\% (152~TWh) are supplied by fossil power generation (hard coal, lignite and gas) in the SLR scenario. This share drops to 28.9\% (141~TWh) in the DLR scenario, while the share of onshore wind power increases from 24\% (117~TWh) to 24.6\% (120~TWh) and of offshore wind power from from 4.2\% (20.6~TWh) to 5.8\% (28.5~TWh). The total generation of the other carriers remain unchanged. We recall that the installed generation capacities are the same for both scenarios. As we show in the following, the reason for this shift in generation lays in an improved exploitation of the wind and transmission infrastructure.

\begin{figure}[htpb]
    \centering
    \includegraphics[width=\linewidth]{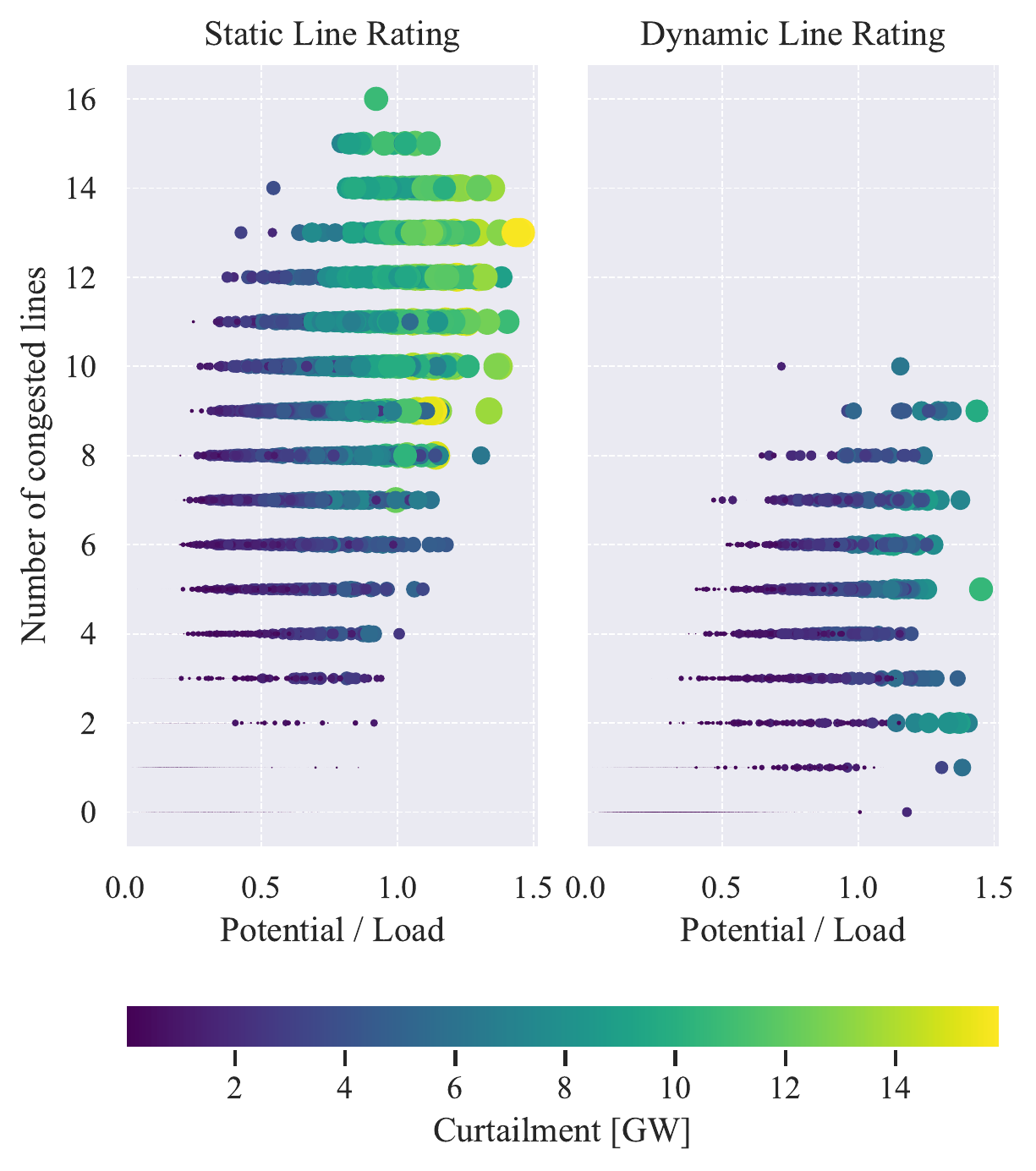}
    \caption{The plot shows the number of congested lines for SLR and DLR as a function of the ratio between the renewable generation potential and the load for each hour during the year. A ratio higher than 1 implies an over-supply which is curtailed. The size and color of the scatter points correspond to the amount of curtailment at the regarded hour.}
    \label{fig:congestion-correlation-2019}
\end{figure}

Fig.~\ref{fig:congestion-correlation-2019} shows the number of congested lines as a function of the total renewable power potential relative to the total load in each time step. The color and size of the dots indicate the renewable curtailment. It stands out that DLR (right panel) leads to a significant decrease in transmission congestion as well as curtailment. While the SLR scenario (left panel) reveals up to 16 congested lines, maximally 10 lines are congested in the the DLR scenario. In particular, at times with a shortage of renewable energy (Potential / Load $<$ 1) the SLR system has to curtail 7 times more power due to transmission congestion. This does not hold for the DLR scenario. The same accounts for times with renewable excess power (Potential / Load $>$ 1), where power is mainly curtailed due to oversupply in the DLR scenario and curtailed due to oversupply and congestion in the SLR scenario.
Solely, from the change in operation, the system saves around 403~mil~Euro of dispatch costs per year throughout the whole simulation year. This translates to 5.5~\% of the total operational expenditure (OPEX). In addition to the OPEX savings, the implementation of DLR leads to a carbon emission reduction of 3.1\% compared to the SLR scenario.

In the 2030 capacity expansion scenario, the impact of DLR becomes even more prominent. Table \ref{tab:installed capacity} illustrates the optimal capacity expansion and installed capacities for the SLR and DLR scenario. In comparison to today's capacity layout, the installed capacity of renewable generators increases by a factor of 2.55 and 2.69, for DLR and SLR respectively. Largely, this increase is driven by the constraint that at least 80\% of the total power generation has to come from renewables. Note that in none of the scenarios, additional gas turbines are added to the system. However, a strong hydrogen infrastructure for power-to-gas-to-power storage is built out, which serves as backup for periods with low renewable potentials.

In comparison to SLR, the system with DLR builds out more offshore wind and less onshore wind, solar and hydrogen infrastructure. In particular, the DLR implementation leverages the offshore wind potentials which, in contrast to the onshore wind and solar power, supply the system with a relatively steady power feed-in. This determining factor also reduces the need for long-term energy storage, embodied by the H$_2$ infrastructure.

\begin{table}
    \caption{Capacity expansion / installed capacity in 2030 study case}
    \begin{tabularx}{\linewidth}{lrrrr}
        {} & \multicolumn{2}{l}{Static Line Rating} & \multicolumn{2}{l}{Dynamic Line Rating} \\
        {} &               Exp.~[GW] &         Inst.~[GW] &                Exp.~[GW] &         Inst.~[GW] \\
        \hline
        Offshore wind & 7.6 & 15.13  & 14.62 & 22.15 \\
        Onshore wind  & 59.72  & 112.91 & 50.27 & 103.46 \\
        Solar   & 118.57 & 167.61 & 105.75 & 154.79 \\
        H$_2$ Fuel cells & 19.07 & 19.07 & 16.4 & 16.4 \\
        \hline
    \end{tabularx}
    \label{tab:installed capacity}
\end{table}

\begin{figure}[htpb]
    \centering
    \includegraphics[width=\linewidth]{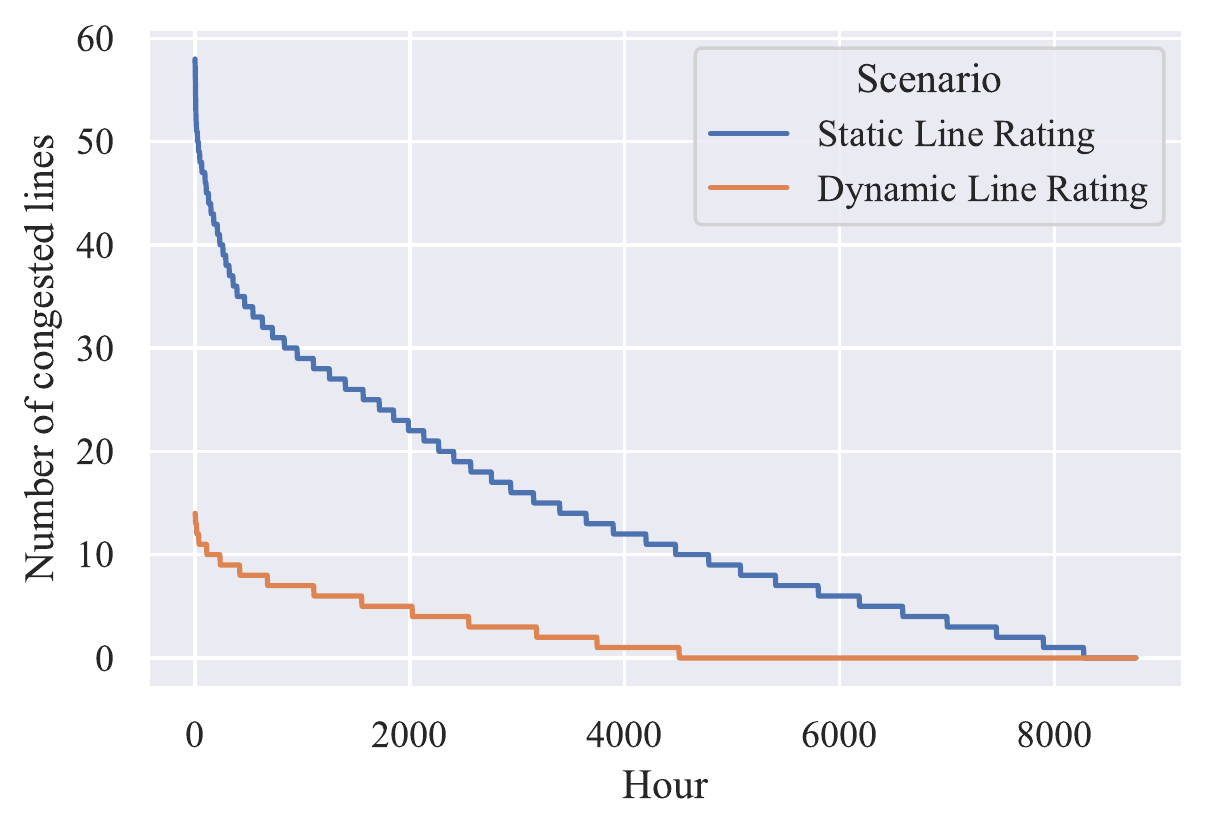}
    \caption{Duration curve for number of congested lines in the 2030 study case.}
    \label{fig:load-duration-2030}
\end{figure}

As depicted in Fig.~\ref{fig:load-duration-2030}, transmission congestion is significantly reduced in the DLR scenario. The figure shows the duration curve of the total congestion throughout the simulation year. In the DLR scenario, approximately half the year is free of congestion while, for SLR, this situation can only be observed for a few hundred hours. Furthermore, the number of maximally congested lines drops from 58 to 14 when going from SLR to DLR.

\begin{figure}[htpb]
    \centering
    \includegraphics[width=\linewidth]{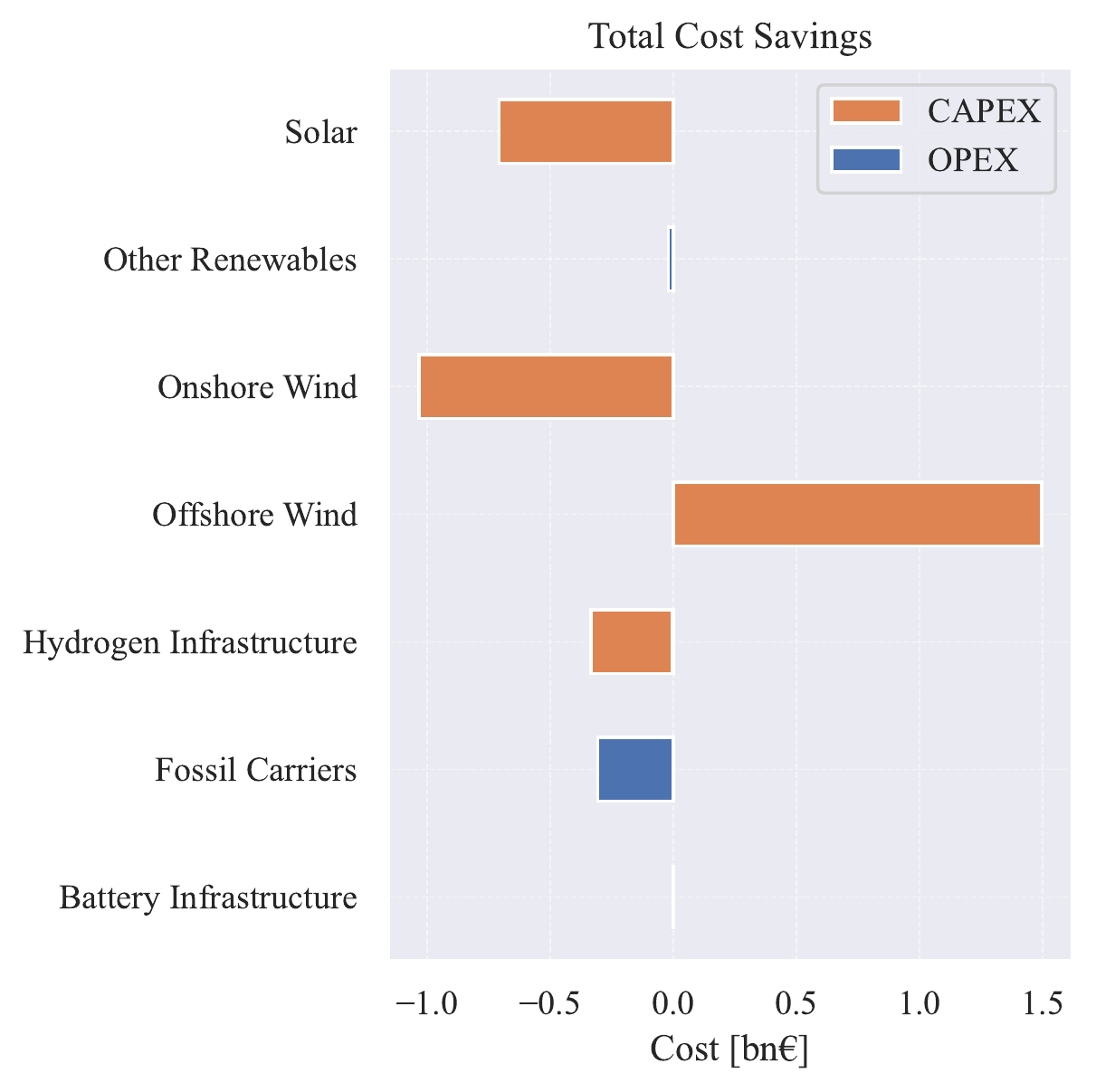}
    \caption{Savings of capital expenditures (CAPEX) and operational expenditures (OPEX) when going from SLR to DLR.  Positive costs represent higher expenses of DLR.}
    \label{fig:cost-bar-2030}
\end{figure}

Fig.~\ref{fig:cost-bar-2030} illustrates the cost savings for the DLR scenario compared to the SLR scenario. Note that positive costs represent higher expenses for the DLR scenario. The capital cost from Fig.~\ref{fig:cost-bar-2030} directly correlates to the capacity expansion given in Table \ref{tab:installed capacity}. We observe significantly less OPEX for fossil energy carriers in the DLR scenario. The total cost savings for the DLR scenario are 3.4\% which corresponds to 908~mil~Euro per year. %\todo[inline]{more positive framing}

\section{Limitations}
\label{sec:limitations}
The presented power system model reveals a high spatial resolution as well as detailed information about installed capacities and renewable potentials. However, it is not able to represent all the power system dynamics. The model is isolated from other countries, neglecting cross-border exchanges. In particular, missing cross-border lines lead to an overestimated curtailment in the north-west of Germany at the substations ``Diele \& Dörpen West'', where much of the curtailed wind power could have been exported to the Netherlands. Further, coal and gas power plants were modelled without ramping limitations.

%\todo{ERA5 underestimates low wind speeds with reference.} % Could not find a reference for this statement. Only that it underestimates wind speeds, but also found one reference saying it overestimates low wind speeds and underestimates high wind speeds (https://rmets.onlinelibrary.wiley.com/doi/full/10.1002/joc.7103).

The optimization of the model is kept linear, meaning that power transmission losses as well as complex power flow dynamics are neglected. Nonetheless, a nonlinear power flow calculation based on the linearly optimized generation dispatch converged for all of the time steps.

\section{Conclusion}
\label{sec:conclusion}
In this paper, we evaluate the effect of DLR on two study cases, namely the German power system in a historical setup for the year 2019 and in a future setup for the year 2030. We show that DLR offers a considerable and feasible complement to transmission capacity expansion. For the 2019 scenario, we observe cost savings of over 403~mil~Euro due to better wind power integration and a shift away from fossil energy carriers. The 2030 study case shows that an implementation of DLR with a parallel expansion of renewable power saves 908~mil~Euro of annual capital and operational system costs. It also reduces the need for hydrogen storage and fuel cells, which must be widely deployed for a renewable energy supply of 80\%  targeted by 2030. This reduction is mainly due to a better integration of offshore wind power and less transmission congestion.

We conclude that in urgent need for decarbonizing the German power system, DLR is a viable complement to the current transmission capacity expansion, not only increasing the total welfare but also reducing the grid congestion.

\newpage

\printbibliography

\appendix
\subsection{Study Cases Parameters}\label{tab:cases_parameters}
\begin{tabularx}{\linewidth}{>{\bfseries}l  X  X}
    \hline
    { }                             &         2019  &                          2030 \\
    \hline
    Generator capacity\\expansion    &            no &                           yes \\
    CO2 Limit                       &  222 mil tons &                  175 mil tons \\
    Base load of nuclear\\and lignite &           yes &                            no \\
    Renewable generation\\constraint &            no &      80\% renewable generation \\
    Generator infrastructure        &          2019 &  2019 still existing in 2030 \\
    Electrical load                 &       605 TWh &                        658 TWh \\
    \hline
\end{tabularx}

\subsection{DLR Factor}\label{dlr_factor}
In the following, we illustrate why the hourly averaged wind speed overestimates the DLR transmission capacity compared to higher resolved data. According to the IEEE standard \cite{IEEEStandardCalculating2012},
\begin{align*}
    P_{max}\propto I_{max} \approxprop \sqrt{\overline{v}_{wind}}
\end{align*}
, where $\overline{v}_{wind}$ denotes the averaged hourly wind speed.
In the following equation,
\begin{align*}
    \sqrt{\overline{v}_{wind}}=\sqrt{\frac{1}{n}\sum_{i}^{n}{v_{wind,i}}} \ge \frac{1}{n}\sum_{i}^{n}{\sqrt{v_{wind_i}}}
\end{align*}
$v_{wind,i}$ corresponds to an sub-hourly wind speed data point going from $i$ to $n$, i.e. from $1$ to $6$ when regarding 10-minute wind speed data. The Equation shows that the root of the averaged sub-hourly wind speed is greater equal the average of the rooted sub-hourly wind speed.

\end{document}